\documentclass[twocolumn,floatfix, nofootinbib,prd,preprintnumbers,showpacs,showkeys,superscriptaddress,longbibliography]{revtex4-2}
\usepackage{slashed}
\usepackage{amsmath}
\usepackage{booktabs}
\usepackage{array}

\usepackage{graphicx} 
\usepackage{orcidlink}
\usepackage[top=2cm,bottom=2cm,left=1cm,right=1cm,marginparwidth=1.75cm]{geometry}
\title{Asy Static Finite Distance}
\usepackage{mathrsfs}
\usepackage{amssymb}
\usepackage{amsmath}
\date{\today}
\usepackage{mathtools}
\usepackage{comment}

\usepackage[mathscr]{euscript}
\usepackage{amsmath}
\usepackage{graphicx}
\usepackage{dcolumn}
\usepackage{bm}
\usepackage{epsfig}
\usepackage{amssymb,latexsym,mathrsfs}
\usepackage{graphicx}
\usepackage{color}
\usepackage{hyperref}
\usepackage{float}
\usepackage{diagbox}

\usepackage{textgreek}

\usepackage{tikz}

\usepackage{amsmath}
\usepackage{physics}
\usepackage{csquotes} 
\newcommand*\diff{\mathop{}\!\mathrm{d}}

\hypersetup{
    colorlinks=true,
    linkcolor=red,
    citecolor=blue,
}

\usepackage{amsmath}
\usepackage{amssymb}
\usepackage{subfigure}
\usepackage{hyperref}
\usepackage{url}
\usepackage{xcolor}
\usepackage{color}
\usepackage{soul}
\definecolor{amaranth}{rgb}{0.9, 0.17, 0.31}
\definecolor{purple(munsell)}{rgb}{0.62, 0.0, 0.77}
\definecolor{americanrose}{rgb}{1.0, 0.01, 0.24}
\definecolor{palatinateblue}{rgb}{0.15, 0.23, 0.89}
\definecolor{royalblue(web)}{rgb}{0.25, 0.41, 0.88}
\definecolor{hanpurple}{rgb}{0.32, 0.09, 0.98}
\definecolor{beaublue}{rgb}{0.74, 0.83, 0.9}
\definecolor{carminered}{rgb}{1.0, 0.0, 0.22}
\definecolor{brightpink}{rgb}{1.0, 0.0, 0.5}
\definecolor{vividviolet}{rgb}{0.62, 0.0, 1.0}
\hypersetup{ linktoc=all,
    colorlinks, linkcolor={palatinateblue},
    citecolor={brightpink}, urlcolor={amaranth}}

\usepackage{comment}

\newcommand{\be}{\begin{equation}}
\newcommand{\ee}{\end{equation}}
\newcommand{\bs}{\begin{split}} 
\newcommand{\bea}{\begin{eqnarray}}
\newcommand{\eea}{\end{eqnarray}}

\newcommand{\ei}[1]{\textcolor{teal}{[{\bf EI}: #1]}} 
\newcommand{\mike}[1]{\textcolor{red}{[{\bf MG}: #1]}} 
 
\newcommand{\ahsan}[1]{\textcolor{blue}{[{\bf AM}: #1]}}

\newcommand{\bes}{\begin{subequations}}
\newcommand{\ees}{\end{subequations}}

\newcommand{\bo}{\raise-1mm\hbox{\Large$\Box$}}

\begin{document}

\preprint{FTPI-MINN-24-24}
\preprint{UMN-TH-4404/24}

\title{There and Back Again: Quantum Radiation from Round-trip Flying Mirrors}
\author{Ahsan Mujtaba\,\orcidlink{0009-0006-2595-6991}}
\email{ahsan.mujtaba@nu.edu.kz}
\affiliation{Physics Department \& Energetic Cosmos Laboratory, Nazarbayev University,\\
Astana 010000, Qazaqstan}
\author{Evgenii Ievlev\,\orcidlink{0000-0002-5935-4706}}
\email{ievle001@umn.edu}
\affiliation{William I. Fine Theoretical Physics Institute, School of Physics and Astronomy,
University of Minnesota,\\
Minneapolis, MN 55455, USA}
\author{Matthew J. Gorban\,\orcidlink{0000-0003-1115-0461}}
\email{matthew\_gorban1@baylor.edu}
\affiliation{Department of Physics, Baylor University,\\
Waco, Texas 76798, USA}
\affiliation{Center for Astrophysics, Space Physics and Engineering Research, Baylor University,\\ Waco, Texas 76798, USA}
\author{Michael R.R. Good\,\orcidlink{0000-0002-0460-1941}}
\email{michael.good@nu.edu.kz}
\affiliation{Physics Department \& Energetic Cosmos Laboratory, Nazarbayev University,\\
Astana 010000, Qazaqstan}
\affiliation{Leung Center for Cosmology and Particle Astrophysics,
National Taiwan University,\\ Taipei 10617, Taiwan}

\begin{abstract} 


Erasing a black hole leaves spacetime flat, so light passing through the region before any star forms and after the black hole’s evaporation shows no time delay, just like a flying mirror that returns to its initial starting point. Quantum radiation from a round-trip flying mirror has not been solved despite the model's mathematical simplicity and physical clarity. Here, we solve the particle creation from worldlines that asymptotically start and stop at the same spot, resulting in interesting spectra and symmetries, including the time dependence of thermal radiance associated with Bose-Einstein and Fermi-Dirac Bogolubov coefficients. Fourier analysis, intrinsically linked to the Bogolubov mechanism, shows that a thermal Bogolubov distribution does not describe the spin-statistics of the quantum field. 
\end{abstract}
\keywords{moving mirrors, acceleration radiation, moving point charge radiation, black hole evaporation}
\pacs{41.60.-m (Radiation by moving charges), 05.70.-a (Thermodynamics),
04.70.Dy (Quantum aspects of black holes),
04.62.+v (Quantum field theory in curved spacetime)}
\maketitle


\section{Introduction}

Investigating the fascinating dynamics of quantum fields through the frequency distribution of their spectra reveals important insight behind particle creation, especially from black holes \cite{Hawking:1974sw}. 
One well-established method to investigate the fundamental mechanisms leading to particle creation from the quantum vacuum is the study of radiation from moving mirrors as an analog to black hole radiance \cite{DeWitt:1975ys, Davies:1976hi,Davies:1977yv}.

To investigate fundamental physical phenomena, moving mirrors provide a simple model for addressing a wide range of complex problems. For example, they have helped understand the information loss problem~\cite{Hotta:2015yla,PisinChen2017,Reyes:2021npy,wilczek1993quantum}, entanglement harvesting \cite{akal2021entanglement,kawabata2021probing,osawa2024final, wald2019particle,Cong:2020nec,Cong:2018vqx}, fluctuation-dissipation~\cite{Hsiang:2024xlh,Xie:2023wvu}, holography \cite{akal2021holographic,akal2022zoo,ievlev2024moving,ling2022reflected,pujolas2008strongly}, conformal field theory \cite{biswas2024moving}, and complexity \cite{sato2022complexity}. 
More generally, the dynamic Casimir effect (DCE) \cite{moore1970quantum}, wherein time-dependent (i.e., moving)  mirrors interact with the quantum vacuum, resulting in particle production (see \cite{dodonov2009dynamical,dodonov2010current,dodonov2020fifty} for reviews), has an expanding literature helping confirm the effectiveness of the moving mirror model \cite{dodonov1996generation,alves2003dynamical,alves2006dynamical,alves2008energy,alves2010exact,good2021quantum}; notably, study of partially reflective mirrors \cite{barton1993quantum,obadia2001notes,nicolaevici2001quantum,haro2008black,nicolaevici2009semitransparency,fosco2017dynamical,Lin:2021bpe}, quantum radiation reaction forces \cite{jackel1992fluctuations, jaekel1993quantum, barton1995quantum1,barton1995quantum2, lambrecht1996motion,jaekel1997movement, alves2008quantum, alves2010quantum, butera2019mechanical, Gorban:2024vss}, and experimental verification \cite{wilson2011observation,lahteenmaki2013dynamical, vezzoli2019optical,paraoanu2020listening,Chen:2020sir,Chen:2015bcg,AnaBHEL:2022sri}. 

\begin{figure}
    \centering
    \includegraphics[width=1\columnwidth]{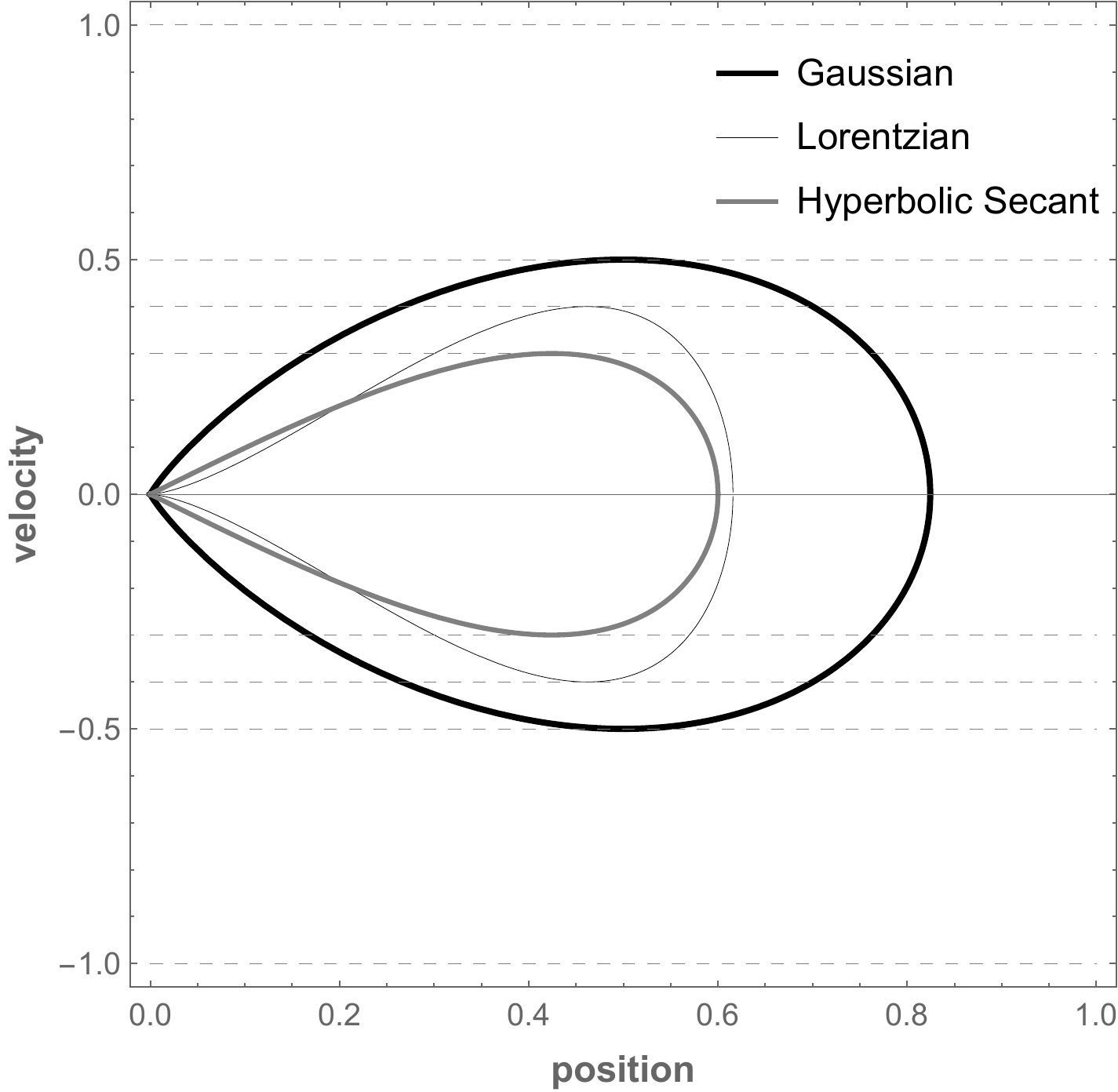}
    \caption{The Gaussian, Lorentzian, and Hyperbolic Secant trajectories, Eq.~(\ref{gausstraj}), Eq.~(\ref{lorentztraj}), and Eq.~(\ref{sechtraj}), are plotted in phase space with maximum velocity  50\%, 40\%, and 30\% the speed of light, respectively.  Closed-path motion requires a finite orbit with a time-asymptotic approach to and from the origin: (position, velocity) $\rightarrow (0,0)$.}  
    \label{GLS}
\end{figure}

A largely unexplored aspect of the moving mirror model is closed-path motion. See Fig. \ref{GLS} for an illustration. Such trajectories asymptotically start and stop at the same spot. Consider the unusual footprint of closed-path trajectories: 
\begin{itemize}
     \item A spectrometer observes no Doppler shift (unlike motion from rest that ends with a constant velocity). 
     \item A light clock observes no time delay (unlike motion that is overall shifted by a distance traveled).
    \item There is no asymptotic evidence that the motion ever happened, except by the particle and energy creation. 
    \end{itemize}
A completely evaporated black hole leaves spacetime flat, with no residual curvature. To test this, one could compare the passage of light through the region before any star formed and after the black hole has fully evaporated; both scenarios should show no time delay. Similarly, a closed-path flying mirror should produce no time delay difference.

Observed particle creation and energy emission are purely due to the mirror's motion rather than its overall position or difference between the initial and end-state. Round-trip solutions have zero net displacements, which allows for understanding how motion alone (i.e., the non-asymptotic, intermediate movement) can influence particle production by quantum fields.

A wide range of round-trip motions can occur, with some exhibiting oscillations that are gradually driven and damped, ultimately leading the object to settle at its equilibrium position, such as in the case of a modulated Gaussian or exponential decay \cite{Gorban:2024vss,Gorban:2023dqz}.  The highly constrained (see Table \ref{accvelpos}) perfectly reflecting round-trip mirror emits the same energy on both sides, unlike net-displaced asymptotically static mirrors. This theoretical symmetry is complemented by experimental potential, which favors a closed path for isolation and study (cf. classical acceleration temperature (CAT) in a box \cite{Mujtaba:2024vqy}). Confined accelerating mirrors could provide clear observational signals that can be analyzed to understand better the external effects of boundary conditions on quantum fields.

Trajectories that asymptotically return to rest with a net positional shift (one-way flights) have been worked out (see `Evanescents' in Table \ref{analogs}).   However, the more restrictive Bogolubov transformation for round-trip trajectories has been intractable. 
This paper presents novel quantum radiative solutions for several closed-path trajectories.  The non-relativistic regime allows analytical solutions of finite particle production from the quantum vacuum.

The structure of the paper is as follows: in section \ref{sec:framework}, we review only the essential formulas for confirming the time and frequency domain analysis of energy production from round-trip flying mirrors. For example, in Eq.~(\ref{betas}), we state the Bogolubov betas are Fourier transforms of the trajectory. Section \ref{sec:GL} (Gauss-Lorentz) is devoted to the spectral analysis of round-trip flying mirror radiation and some exciting spectra.  Leveraging the insight gained from these examples, we compute round-trip trajectories that exhibit thermal Bogolubov distributions in section \ref{sec:BE} (Bose-Einstein) and section \ref{sec:FD} (Fermi-Dirac). Finally, section~\ref{sec:conclusions} summarizes the main findings.  Units are $c = \mu_0 = \epsilon_0 =1$, where Planck's constant $\hbar$ is left unset to emphasize the quantum field effects.



\section{Framework}
\label{sec:framework}
Through time domain analysis---considering both sides of a non-relativistic moving mirror---one can determine the magnitude of the quantum reaction force (e.g., \cite{Ford:1982ct}),
\be F = -\frac{\hbar}{6\pi}\frac{\diff{\alpha(\tau)}}{\diff{\tau}} \quad \xrightarrow{\textrm{non-rel.}} \quad F = -\frac{\hbar}{6\pi}\frac{\diff{a(t)}}{\diff{t}},\label{reaction}\ee
and the quantum power radiated by the motion (e.g., \cite{Zhakenuly:2021pfm}),
\be P =  \frac{\hbar \alpha^2(\tau)}{6\pi} \quad \xrightarrow{\textrm{non-rel.}} \quad P = \frac{\hbar a^2(t)}{6\pi}.\label{power}\ee
Here $\alpha(\tau)$ is the proper acceleration, $\tau$ is the proper time, $a(t)$ is the coordinate acceleration, and $t$ is the coordinate time. The total energy emitted may be found via time analysis, using either the quantum reaction force or the radiated quantum power (see, e.g., the classical analog via Eq. 9.1.2 of \cite{Feynman:1996kb}):
\be E = \int_{-\infty}^{+\infty} P(t) \diff{t} = \int_{-\infty}^{+\infty} F(t) v(t) \diff{t}.\label{energyfrompower}\ee 

The total energy emitted can also be derived via spectral analysis in the frequency domain. 
The beta Bogolubov coefficients relevant for both sides of the non-relativistic mirror trajectory are given by
(see Appendices~\ref{sec:FB_via_mapping} and \ref{sec:FB_via_mirror})
\be |\beta_{pq}|^2 \approx \frac{4}{\pi} p q |z_\omega|^2, \label{betas}\ee
where the Fourier transform of $z(t)$ is $z_\omega = \mathcal{F}z_t$ (see Eq.~\eqref{fourier_def} for conventions) and $\omega = p+q$.  Here, $q$ and $p$ are the two in-out vacuum state frequency modes, respectively, e.g., \cite{good2013time}. The particle spectrum in terms of frequency $p$ may then be found by integrating over frequency $q$, (see e.g., \cite{Birrell:1982ix})
\be N(p) = \int_0^\infty |\beta_{pq}|^2 \diff{q},\label{particlespectrum}\ee
while the total particle count is found by integrating the Bogolubov coefficients over both $q$ and $p$, (see e.g., \cite{Fabbri})
\be N = \int_0^\infty\int_0^\infty |\beta_{pq}|^2 \diff{q}\diff{p}.\label{totalparticles}\ee
The total energy emitted is found by associating a quantum of energy $\hbar p$ with the particle distribution, (see e.g., \cite{walker1985particle})
\be E =  \int_0^\infty \hbar p \; N(p) \diff{p}.\label{energyfrombetas}\ee
We restrict our scope to causal time-like trajectories with velocities,
\be |\dot{z}(t)| < \textrm{speed of light},\ee
travelling along rectilinear paths starting at rest and returning to their original position:
\be \textrm{round-trip:} \qquad \lim_{t\to\pm \infty} z(t) = 0, \label{loop}\ee
where we have taken the liberty to choose the origin as the initial and final resting place. 

\begin{center}
\begin{table}[H]
\centering
\begin{tabular}{>{\raggedright\arraybackslash}p{2cm}cccc}
\toprule
 & \textbf{Constraint} & \textbf{Feature} & \textbf{Example}  \\
\midrule
Acceleration & $\ddot{z} \to 0$ & Sub-luminal & Carlitz-Willey \cite{carlitz1987reflections}  \\
Velocity & $\dot{z} \to 0$ & IR-Finite & Walker-Davies \cite{Walker_1982}  \\
Position & $z \to 0$ & Round-Trip & Gauss-Lorentz [Sec.~\ref{sec:GL}]  \\
\bottomrule
\end{tabular}
\caption{Viewed in terms of coordinate time derivative constraints, the physical features like sub-luminal speeds or IR-finite spectra suggest round-trip trajectories may offer additional physically desirable characteristics.  Indeed, the additional symmetry of a closed path means identical energy emission on both sides of the mirror, e.g.,  \cite{Good:2017kjr}. The arrows in the constraint column are asymptotic coordinate time limits, $t \to \pm\infty$.}
\label{accvelpos}
\end{table}
\end{center}

\begin{center}
\begin{table}[H]
\centering
\begin{tabular}{>{\raggedright\arraybackslash}p{3.5cm}ccccc}
\toprule
\textbf{Mirror Analogs} & \textbf{Energy} & \textbf{Rapidity} & \textbf{Particles} & \textbf{Orbit} \\
\midrule
Cosmologies\;\;\cite{Good:2020byh,Fernandez-Silvestre:2021ghq,Fernandez-Silvestre:2022gqn} & $\infty$ & $\infty$ & $\infty$ & $\infty$ \\
Black holes\;\; \cite{Good:2016oey,good2020particle,Good:2020fjz,Foo:2020xmy,Foo:2020bmv} & $\infty$ & $\infty$ & $\infty$ & $\infty$ \\
Extremals\;\;\;\; \cite{Liberati:2000sq,good2020extreme,Foo:2020bmv,Rothman:2000mm,Kumar:2023kse} & \checkmark & $\infty$ & $\infty$ & $\infty$ \\
Remnants \;\;\;\;\cite{Good:2016atu,Good:2018ell,Good:2018zmx,Myrzakul:2018bhy,Good_2015BirthCry,Good:2016yht} & \checkmark & \checkmark & $\infty$ & $\infty$ \\
Evanescents\; \cite{ GoodMPLA,Good:2017ddq,Good:2018aer,Good:2019tnf} & \checkmark & \checkmark & \checkmark & $\infty$ \\
\midrule
\textit{Round-Trip} & \checkmark & \checkmark & \checkmark & \checkmark \\
\bottomrule
\end{tabular}
\caption{The round-trip mirror solutions in this paper are unique as measured against previously studied trajectories. In this Table, `Cosmologies' refers to dS, AdS, and SdS spacetimes, `Black holes' refer to Schwarzschild, RN, and Kerr, and `Extremals' refer to the extremal RN/Kerr. `Remnants' refers to asymptotically non-zero constant velocity trajectories with a residual field mode Doppler shift. The label `Evanescents' describes finite-particle evaporation trajectories that are asymptotically inertial with zero-velocity (no residual field mode Doppler shift) yet do not return to their origin (non-zero net displacement), whose footprint can be measured by a pair of two asymptotic light-rays delayed in time. In 1+1 dimensions, an `orbit' refers to the mirror's time-integrated displacement along its rectilinear path, $\int z(t) \diff{t}$, which can be finite if it returns to the same position (one full orbit) and infinite if it does not. In the limit that $t\to \pm \infty$, we have $z(t) \to 0$, and $z(t) \nrightarrow 0$, for finite and infinite orbits, respectively.}
\label{analogs}
\end{table}
\end{center}

\section{Gauss-Lorentz}
\label{sec:GL}
Several well-known functions satisfy the condition of returning to rest at the origin, Eq.~(\ref{loop}), 
and whose Fourier transform may be computed to find the Bogoluvbov coefficients, Eq.~(\ref{betas}). Since we require the trajectory to be sub-luminal, $|\dot{z}| < 1$ [and in this paper, the mirrors do not just move to the left but also move to the right, so $\dot{z}$ is sign-indefinite, and the absolute value sign is essential to convey this non-monotonicity], it is natural to construct a physically reasonable $z(t)$ in terms of the maximum speed, 
$v = \max(|\dot{z}|) <1$.
This section briefly examines the analytical solutions for the Gaussian, Lorentzian, and other interesting motions.  
\subsection{Gauss}
The Gaussian flying mirror takes the form (recalling the maximum velocity $v$ is dimensionless and the acceleration scale $\kappa$ is dimensionful):
\be z(t) = \frac{v}{\kappa} \exp \left(\frac{1}{2}-\frac{1}{2} (\kappa t)^2\right).\label{gausstraj}\ee
The beta coefficients in the non-relativistic limit, Eq.~(\ref{betas}), are of course, also Gaussian, 
\be |\beta_{pq}|^2 = \frac{4 p q v^2}{\pi  \kappa ^4} \exp \left(1-\frac{(p+q)^2}{\kappa ^2}\right).\ee
The particle spectrum, Eq.~(\ref{particlespectrum}), is then
\be N(p) = \frac{2 p v^2 e^{1-\frac{p^2}{\kappa ^2}}}{\pi  \kappa ^2}-\frac{2 e p^2 v^2 \text{erfc}\left(\frac{p}{\kappa }\right)}{\sqrt{\pi } \kappa ^3}. \label{GaussianParticles}\ee
The total energy, Eq.~(\ref{energyfrombetas}), and total amount of particles created, Eq.~(\ref{totalparticles}), are:
\be E = \frac{e \hbar \kappa v^2}{8 \sqrt{\pi }}, \qquad N = \frac{e v^2}{3 \pi } = 0.288419 v^2. \label{GAUSS_n_PARTICLE EQ}\ee
The energy above can be checked in the time domain against the radiation reaction, Eq.~(\ref{reaction}), and power, Eq.~(\ref{power}), using Eq.~(\ref{energyfrompower}). 
With its smooth, symmetric bell-shaped curve, the Gaussian trajectory concentrates particle production around the peak, offering a well-focused radiation signature.
\subsection{Lorentz}
Consider the Lorentzian flying mirror, uniquely constructed so that the max velocity $v < 1$:
\be z(t) = \frac{v}{\kappa}\frac{8}{3 \sqrt{3}  \left(\kappa ^2 t^2+1\right)}.\label{lorentztraj}\ee
The beta coefficients, Eq.~(\ref{betas}), have a Fourier transform that hints at exponential attenuation in particle production: 
\be |\beta_{pq}|^2 =\frac{128 p q v^2}{27 \kappa^4} e^{-\frac{2 (p+q)}{\kappa }}.\ee
Indeed, the particle spectrum, Eq.~(\ref{particlespectrum}), is more simple than the Gaussian of Eq.~(\ref{GaussianParticles}), possessing a linear coupling to exponential decay 
\be N(p) = \frac{32 p v^2}{27\kappa^2} e^{-\frac{2 p}{\kappa }}.\ee
The total energy, Eq.~(\ref{energyfrombetas}), and total amount of particles created, Eq.~(\ref{totalparticles}), are coincidently similar ($E/N = \hbar \kappa$):
\be E = \frac{8 v^2}{27} \hbar \kappa , \qquad N = \frac{8 v^2}{27} = 0.296296 v^2.\label{lorentz_n_particle_eq}\ee
Like the Gaussian, the energy obtained in the frequency domain is consistent with the time domain using the radiation reaction and power, i.e., Eq.~(\ref{energyfrompower}). 

\begin{figure}
    \centering
    \includegraphics[width=\columnwidth]{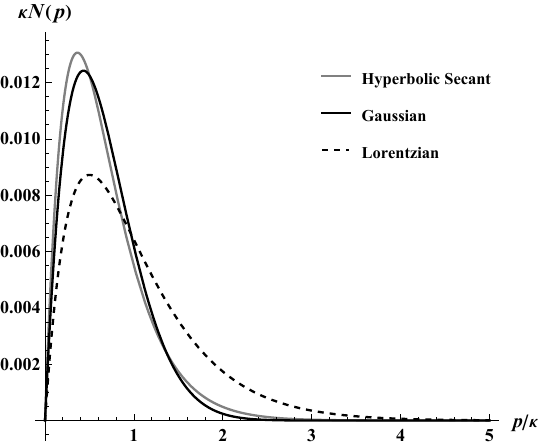}
    \caption{The figure compares the Gaussian, Lorentzian, and Hyperbolic Secant mirrors' particle spectrum with maximum velocities set at $v=0.2$. It can be seen that Gaussian and Sech share similar particle counts, $N$,  which is also depicted through Eq. (\ref{GaussianParticles}) and Eq. (\ref{HS_n_PARTICLES_eq}). The Lorentz curve, Eq. (\ref{lorentz_n_particle_eq}), is less peaked but has a broader spectrum with particle count, $N$ close to the Sech trajectory total emission.}
    \label{hyp_gauss_lor_spectr-fig} 
\end{figure}

However, unlike the Gaussian, the Lorentizian's characteristic long tails allow for a slower UV decay, resulting in broader particle production over a wider range of times and frequencies.
See Fig.~\ref{hyp_gauss_lor_spectr-fig} for an illustration.

\subsection{Hyperbolic Secant}
Let us consider the hyperbolic secant flying mirror with max velocity $v<1$:
\be z(t) = \frac{2 v}{\kappa} \text{sech}(\kappa  t).\label{sechtraj}\ee
The beta coefficients, Eq.~(\ref{betas}), are of course, also hyperbolic secants:
\be |\beta_{pq}|^2 =\frac{8 p q v^2}{\kappa^4} \text{sech}^2\left(\frac{\pi  (p+q)}{2 \kappa }\right).\ee
However, the particle spectrum, Eq.~(\ref{particlespectrum}), is a linear-log scaling, 
\be N(p) = \frac{32 p v^2}{\pi^2 \kappa^2} \ln \left(e^{-\frac{\pi  p}{\kappa }}+1\right).\ee
The total emission,
\be E = \frac{28 v^2}{45 \pi } \hbar \kappa, \qquad N = \frac{24 v^2 \zeta(3)}{\pi^4} = 0.296167 v^2, \label{HS_n_PARTICLES_eq}\ee
has a particle count which is coincidently close to the Lorentzian particle count, despite its steeper decay in both time and frequency domains.  Interestingly, UV-suppressed particle production is almost perfectly balanced by IR-increased particle production, leading to a more sharply defined emission profile than the broader Lorentzian.


\subsection{Oscillations, UV-cuts, and $\mathcal{T}$-asymmetry}

Round-trip trajectories may be solved with interesting particle spectra whose relationship to maximum velocity is less arithmetically simple.  For instance, using a linear function of the maximum velocity, $J(v)$ (different for each unique trajectory), facilitates solving a Quad-Lorentz-type trajectory
\be z(t) = \frac{J(v)}{\kappa}\frac{1}{(\kappa t)^2 + (\kappa t)^{-2}},\label{oscillatingtraj}\ee
which has an oscillating particle spectrum \be N(p) = \frac{J(v)^2 p}{4\kappa^2} e^{-\frac{\sqrt{2} p}{\kappa }} \left(2 -\cos \frac{\sqrt{2} p}{\kappa }\right)\label{quad-lorentz spectra_eq},\ee
with total particle count
\be N = \frac{J(v) ^2}{4} = 0.354852 v^2. \ee 
Or consider the Sinc trajectory, 
\be z(t) = \frac{J(v)}{\kappa} \frac{\sin (\kappa  t)}{\kappa  t},\label{sinc}\ee
which has a UV-finite cut-off in its particle spectrum,
\be N(p) = \frac{J(v)^2 p}{\kappa^4} (\kappa-p)^2, \quad \to \quad N = \frac{J(v) ^2}{12} = 0.438009 v^2,\label{cutoff}\ee
for $\kappa > p$ and $N(p) = 0$ otherwise. Likewise, the Jinc (Sombrero) trajectory,
\be z(t) = \frac{J(v)}{\kappa} \frac{J_1(\kappa t)}{\kappa t},\label{jinctraj}\ee
may also be used to compute the particle spectrum given by:
\be
N(p)=\frac{2 J(v)^{2}p}{2\pi^{2}\kappa^{6}} (3\kappa+p)(\kappa-p)^{3} ,
\label{jinc_N(p)_equation}
\ee
which shows the UV-finite-cutoff for $\kappa>p$ and $N(p)=0$


The smooth, asymptotic high-frequency behavior typically seen in acceleration radiation contrasts with Eq.~(\ref{cutoff}). Oscillating worldlines, like Eq.~(\ref{sinc}) and Eq.~(\ref{jinctraj}), influence the quantum field by imposing a maximum frequency limit (in the non-relativistic regime). Similarly, Hawking radiation should exhibit an upper-frequency limit dictated by the black hole's total mass, beyond which photon emission is forbidden by energy conservation. Eq.~(\ref{cutoff}) [and spectra like it] may serve as a helpful analog for a sharp cutoff in the non-thermal emission phase of a black hole.

\begin{figure}
    \centering
    \includegraphics[width=1\linewidth]{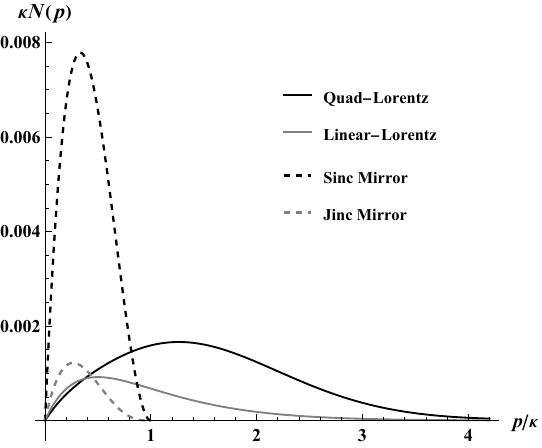}
    \caption{The figure compares the quantum particle spectra of the Quad-Lorentz, Eq. (\ref{quad-lorentz spectra_eq}), Linear-Lorentz, Eq. (\ref{linear-lorentz particlespectra_eq}), Sinc, Eq. (\ref{cutoff}), and Jinc, Eq. (\ref{jinc_N(p)_equation}), mirrors. The Sinc and Jinc spectra possess an interesting non-relativistic UV-cutoff $p/\kappa\leq1$ while the Lorentz spectra asymptotically decay with increasing frequency. }
    \label{fig:quadlinearsincjincspectra}
\end{figure}

In addition to UV-cutoffs, time-asymmetric round-trip trajectories, $z(-t) = -z(t)$, are also possible, e.g., a Linear-Lorentz type trajectory, 
\be z(t) = \frac{v t}{\kappa ^2 t^2+1},\label{timeoddtraj}\ee
which has spectral decay similar to the Lorentzian,
\be N(p) = \frac{v^2 p}{2\kappa^2} e^{-\frac{2 p}{\kappa }}, \quad \to \quad N = \frac{v^2}{8}. \label{linear-lorentz particlespectra_eq}\ee
 A comparison between the Linear-Lorentz spectrum, Eq. (\ref{linear-lorentz particlespectra_eq}) and Quad-Lorentz spectrum Eq. (\ref{quad-lorentz spectra_eq}) is shown in Fig. (\ref{fig:quadlinearsincjincspectra}). These examples demonstrate a diverse range of possible particle spectra. The following sections will use this observation to derive trajectories that radiate thermal Bogolubov distributions.


\section{Bose-Einstein}
\label{sec:BE}
Consider a new worldline, 
\be z(t) = \frac{J(v)}{\kappa} \frac{W(e^{\kappa t})}{(1+W(e^{\kappa t}))^3},\label{BEtraj}\ee
which satisfies the closed-path criteria, Eq.~(\ref{loop}).  To compute the beta coefficients, we present the formula,
\be \left|\mathcal{F}\left[ \partial^n_t f(t)\right] \right|^2 = \frac{\omega^{2n -3}}{e^{2\pi \omega}-1},\quad \textrm{where}, \quad f(t) = W(e^t),\label{FTbose}\ee
where $n$ is the number of derivatives.  
This formula can be easily derived using the results of \cite{Ievlev:2023inj} and the Fourier-of-derivative property Eq.~\eqref{fourier_of_derivative}.
Using our case, $n=2$, the beta coefficients are:
\be |\beta_{pq}|^2 =\frac{4 J(v)^2 p}{\pi \kappa^5} \frac{q (p+q)}{e^{2 \pi  (p+q)/\kappa}-1}, \quad \xrightarrow{q\gg p} \quad \frac{4 J(v)^2 p}{\pi \kappa^5}\frac{q^2}{e^{2 \pi  q / \kappa}-1}. \label{3Dbetamirror}\ee
Notice the 3D Planck distribution form in frequency $q$, which results after applying Hawking's high-frequency approximation \cite{Hawking:1974sw}. The exact particle spectrum can be analytically found by integrating over $q$,
\be N(p) = \frac{J(v)^2 p^2}{\pi^3\kappa^3} \text{Li}_2\left(e^{-2 \pi  p/\kappa}\right)+\frac{J(v)^2 p}{\pi^4\kappa^2} \text{Li}_3\left(e^{-2 \pi  p/\kappa}\right).\label{polylogBE} \ee
The total energy and total particle count is 
\be E = \frac{J(v)^2}{1512\pi} \hbar \kappa, \qquad N = \frac{\zeta (5)}{2 \pi ^6} J(v)^2 = 0.196763 v^2,\ee
where the last step uses an analytically lengthy but straightforward $J(v)$ as a function of the maximum velocity, $v<1$. 

These computations demonstrate that a high-frequency 3D Planck form for a beta coefficient squared, Eq.~(\ref{3Dbetamirror}), does not necessarily result in a standard 3D Planck form for the particle spectrum, 
Thus, assigning a temperature based on the Planckian form of the beta coefficients in Eq.~(\ref{3Dbetamirror}) is equivalent to assigning a temperature to the polylog form of Eq.~(\ref{polylogBE}). 

The Planck distribution derived from the Fourier transform modulus squared, Eq.~(\ref{FTbose}), is an essential physical result because it directly demonstrates that thermal radiation, if characterized by the quantum Bogolubov mechanism, can emerge from classical trajectories (e.g., Eq.~(\ref{BEtraj}) product logs). Assigning a temperature to the spectrum, Eq.~(\ref{3Dbetamirror}), [assuming the appearance of the Planck factor in the Bogolubov coefficients necessarily indicates thermality] establishes a connection between the system's classical dynamical behavior (i.e., the mirror's motion) and the thermal properties of the quantum field.

\section{Fermi-Dirac}
\label{sec:FD}
\begin{figure}
    \centering
    \includegraphics[width=1\columnwidth,height=8cm]{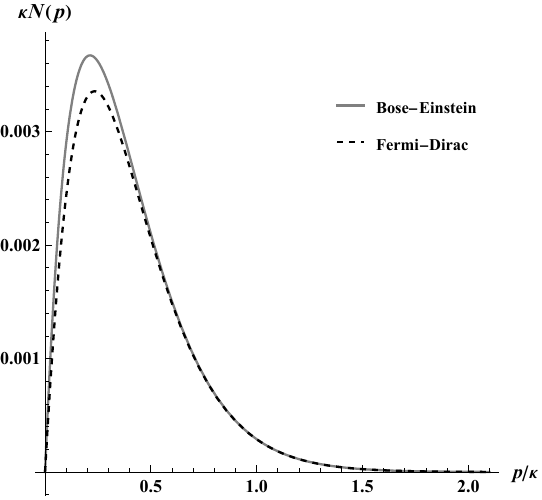}
    \caption{The figure shows the comparison between the particle spectra Eq.~(\ref{polylogBE}) and Eq.~(\ref{polylogFD}) of the Bose-Einstein and Fermi-Dirac distributions respectively. Despite the quantum radiation being from the 1D moving mirror model, the distributions behave like 3D Planck curves with associated temperatures, graybody factors, and no IR divergences.}
    \label{fig:bose-fermi comparison spectra}
\end{figure}

The Bose-Einstein form for the beta coefficients, Eq.~(\ref{3Dbetamirror}), is not the only thermal Bogolubov distribution possible. Consider the following round-trip trajectory,
\be z(t) = \frac{J(v)}{\kappa}\frac{\left(1-W\left(e^{\kappa t}\right)\right) \sqrt{W\left(e^{\kappa t}\right)}}{2 \left(W\left(e^{\kappa t}\right)+1\right)^3}.\label{FDtraj}\ee
One can show that a formula similar to Eq.~(\ref{FTbose}) holds (see \cite{Ievlev:2023akh} and Eq.~\eqref{fourier_of_derivative} here),
\be \left|\mathcal{F}\left[ \partial^n_t g(t)\right] \right|^2 = \frac{\omega^{2n -3}}{e^{2\pi \omega}+1},\quad \textrm{where}, \quad g(t) = 2\sqrt{W(e^t)},\label{FTfermi}\ee
where again $n$ is the number of derivatives, and in our case $n=2$. The beta coefficients are the same as Bose-Einstein beta coefficients, Eq.~(\ref{3Dbetamirror}),
\be |\beta_{pq}|^2 =\frac{4 J(v)^2 p}{\pi \kappa^5} \frac{q (p+q)}{e^{2 \pi  (p+q)/\kappa}+1}, \quad \overset{q\gg p}{\rightarrow} \quad \frac{4 J(v)^2 p}{\pi \kappa^5}\frac{q^2}{e^{2 \pi  q / \kappa}+1}, \label{3DbetamirrorFD}\ee
except these have Fermi-Dirac (FD) form. The particle spectrum is similar, see Fig. \ref{fig:bose-fermi comparison spectra}, to Eq.~(\ref{polylogBE}),
\be N(p) = -\frac{J(v)^2 p^2}{\pi^3\kappa^3} \text{Li}_2\left(-e^{-2 \pi  p/\kappa}\right)-\frac{J(v)^2 p}{\pi^4\kappa^2} \text{Li}_3\left(-e^{-2 \pi  p/\kappa}\right), \label{polylogFD} \ee
but differs by an overall sign and negative signs in the polylog arguments. 
These results complement works that discuss FD-distributed spin-0 particles \cite{Haro:2008zza, Nicolaevici:2009zz,Elizalde:2010zza,Nikishov:1995qs,Takagi:1986kn}, demonstrating the critical fact that beta Bogolubov coefficients may appear thermal via a Planck distribution when the particle spectrum, $N(p)$, is not of canonical Planck-form.  

Although the Bogolubov coefficients for bosons can exhibit a Fermi-Dirac distribution, they do not accurately reflect the spin-0 particle statistics of the emitted scalars. This result challenges the naive assumption that the appearance of a Planck factor in the Bogolubov coefficients necessarily indicates thermality [as is tempting to do from the results given in the previous section]. The scenario changes when considering infinite particle count and energy emission cases, such as during asymptotic infinite proper acceleration toward a light-cone asymptote, e.g., \cite{Chen:2017lum}. In this situation, a Planck factor in the modulus of the Bogolubov coefficient accurately describes thermal radiation from a Schwarzschild black hole, a widely accepted result, see, e.g., \cite{Fulling_optics}. Our result demonstrates a thermal Bogolubov distribution does not describe the spin-statistics of a finite orbit quantum field.

\section{Conclusions}
\label{sec:conclusions}
We have provided a novel approach for solving the problem of quantum radiation from vacuum, applying Fourier analysis to the Bogolubov mechanism. The Fourier-Bogolubov method for determining quantum radiation spectra is straightforward, and its application yields remarkably diverse and elementary results. Using non-relativistic moving mirrors, see Table \ref{summarytable}; we have analytically solved the production of quantum particles in several canonical motions that retrace to their original starting points, such as the Gaussian and Lorentzian trajectories. 

Newfound worldlines emit bosons with Bose-Einstein and, surprisingly, Fermi-Dirac beta Bogolubov distributions, explicitly demonstrating that a thermal Bogolubov distribution cannot characterize the spin-statistics of a quantum field with a finite orbit. Concise analytic solutions to closed-path trajectories reveal the relationship between time-dependent motion and frequency-dependent particle production.  The Fourier-Bogolubov connection provides a physically intuitive and powerful technique for investigating tailored particle spectra from the quantum vacuum.

\begin{center}
\begin{table}[H]
\centering
\begin{tabular}{>{\raggedright\arraybackslash}p{2.3cm}cccc}
\toprule
 \textbf{Trajectory} & \textbf{Worldline} & \textbf{Particles} & \textbf{Energy} &   \\
Function & $z(t)$ & $N/v^2$ & $E/\hbar\kappa v^2$ &   \\
\midrule
Gauss & Eq.~(\ref{gausstraj}) & $0.288419$ & $0.191703$ &  \\
Lorentz & Eq.~(\ref{lorentztraj}) & $0.296296$ & $0.296296$ &  \\
Sech & Eq.~(\ref{sechtraj}) & $0.296167$ & $0.198059$ &  \\
Sinc & Eq.~(\ref{sinc}) & $0.438009$ & $0.175204$ &  \\
Jinc & Eq.~(\ref{jinctraj}) &  $0.059173$ &  $0.020288$ &  \\
Quad-Lorentz & Eq.~(\ref{oscillatingtraj}) &  $0.354852$ &  $0.564566$ &  \\
Linear-Lorentz & Eq.~(\ref{timeoddtraj}) & $0.125000$ & $0.125000$ &  \\
Bose-Einstein & Eq.~(\ref{BEtraj}) & $0.196763$ &  $0.076811$ &  \\
Fermi-Dirac & Eq.~(\ref{FDtraj}) &  $0.183985$ &  $0.074217$ &  \\
\bottomrule
\end{tabular}
\caption{A summary of the round-trip solutions for easy reference to the trajectories in-text. Here, the particle count is the ratio $N/v^2$, and the energy emission is the ratio $E/\hbar\kappa v^2$. The Lorentz and Linear-Lorentz identical particle-to-energy ratios and the similar particle count between Lorentz and Sech are coincidences resulting from the non-relativistic approximation. }\label{summarytable}
\end{table}
\end{center}


\section{Appendix}
This Appendix includes some functional relations and finer details that clarify the non-relativistic UV-cutoff quantum spectra in \ref{appendix:sinc} (sinc) and \ref{appendix:jinc} (jinc), derives total particle production confirming the Bogolubov method in \ref{appendix:N(w)}, and provides two methods for deriving Eq.~(\ref{betas}) in \ref{sec:FB_via_mapping} (electron) and  \ref{sec:FB_via_mirror} (mirror). 



\subsection{Derivation of UV-cutoff Spectrum from Sinc Mirror Trajectory; Eq.~(\ref{cutoff}) from Eq.~(\ref{sinc})} \label{appendix:sinc}
The sinc trajectory,  Eq.~(\ref{sinc}), is given as:
\be
z(t)=\frac{J(v)}{\kappa}\frac{\sin(\kappa t)}{\kappa t}
\label{sinc_traj_eq_appendix}
\ee
The Fourier transform of the Eq. (\ref{sinc_traj_eq_appendix}) is given as:
$$\mathcal{F} z(t)=\frac{1}{\sqrt{2\pi}}\int_{-\infty} ^\infty z(t) e^{-i\omega t} dt,$$
which converts $z(t)$ to the frequency domain as:
\be
z(\omega)=\frac{J(v)\sqrt{\pi}}{2\sqrt{2}}(1+\textrm{sgn}(\kappa-\omega)),
\label{sinc_freq_traj_eq}
\ee
where sgn is piece-wise:
$$
\text{sgn}(\kappa-\omega) =
\begin{cases} 
1 & \text{if } \kappa>\omega, \\
0 & \text{if } \kappa=\omega, \\
-1 & \text{if } \kappa < \omega.
\end{cases}
$$
The sinc trajectory in frequency space, Eq. (\ref{sinc_freq_traj_eq}), has Bogolubov coefficients that be calculated using Eq. (\ref{betas}), remembering that $\omega=p+q$:
\be
|\beta_{pq}|^{2}=\frac{J(v)^{2} pq}{2\kappa^{4}}(1+\text{sgn}(\kappa-(p+q)).
\label{appendix_betasinc}
\ee
Eq. (\ref{appendix_betasinc}) can be used to derive the particle spectrum using Eq. (\ref{particlespectrum}), which gives Eq.~(\ref{cutoff}):
\be
N(p) = \int_0 ^\infty |\beta_{pq}|^2 \diff{q} =\frac{p J(v)^{2}}{\kappa^4}(\kappa-p)^2.
\label{appendix_sinc_N(p)_eq}
\ee
To obtain the total particle count, $N$, we need only integrate over $p\leq \kappa$. $N$ becomes:
\be
N=\int_0 ^\kappa \frac{p J(v)^{2}}{\kappa^4}(\kappa-p)^2 \diff{p} = \frac{J(v)^2}{12} = 0.438009 v^2.
\ee
The energy for the sinc mirror can also be found integrating Eq. (\ref{appendix_sinc_N(p)_eq}) as follows:
\be
E_{\textrm{sinc}}=\int_0 ^\kappa \hbar p \; N(p) \diff{p} = \int_0 ^\kappa \hbar p \frac{p J(v)^{2}}{\kappa^4}(\kappa-p)^2 \diff{p}, \label{appendix_energy_sinc_eq}
\ee
yielding a total energy emitted,
\be E =\frac{\hbar \kappa J(v)^2}{30} = 0.175205 \hbar \kappa v^2,\ee
which can be confirmed via time-analysis of the power and self-force using Eq. (\ref{energyfrompower}).

\subsection{Derivation of UV-cutoff Spectrum from Jinc Mirror Trajectory; Eq.~(\ref{jinctraj})} \label{appendix:jinc}
The Jinc trajectory is given by:
\be
z(t)=\frac{J(v)}{\kappa}\frac{J_{1}(\kappa t)}{\kappa t},
\label{jinc_traj_append}
\ee
where $J_1(\kappa t)$ is the Bessel function of the first kind, and $J(v)$ is our convenient function that depends linearly on the maximum speed, $v$.
We can always compute the Fourier transform of Eq. (\ref{jinc_traj_append}) by using the standard Fourier transform mentioned in Eq. (\ref{fourier_def}), giving us:
\be
z_{\omega} =
\begin{cases} 
\frac{J(v) \sqrt{2 (\kappa^{2}-{\omega}^{2})}}{\sqrt \pi\kappa^{3}} & \text{if } \kappa>\omega, \\
0 & \text{if } \kappa=\omega, 
\end{cases} \label{fourier_jinc_append}
\ee
showing an unusual UV-cutoff for frequencies $\omega>\kappa$ in the expression for $\mathcal{F}z_{\omega}$. 
Eq. (\ref{fourier_jinc_append}) is used for $\kappa>\omega$ to compute the Beta coefficients for the Jinc (Sombrero) mirror by Eq. (\ref{betas}) for $\omega=p+q$ as:
\be
|\beta_{pq}|^{2}=\frac{8 J(v)^{2} pq(\kappa^{2}-(p+q)^{2})}{\pi^{2}\kappa^{6}},
\label{jinc-betasquared_append}
\ee
By using Eq. (\ref{particlespectrum}) over the specified limits of $q$, for Eq. (\ref{jinc-betasquared_append}) we get
\be
N(p)=\int_{0}^{\kappa-p}|\beta_{pq}|^{2} dq=
\frac{2 J(v)^{2}p}{2\pi^{2}\kappa^{6}} (3\kappa+p)(\kappa-p)^{3}.
\label{jinc_particle_spectrum_equat_append}
\ee
Eq. (\ref{jinc_particle_spectrum_equat_append}) can lead to the total number $N$ of particles for Jinc by using Eq. (\ref{totalparticles}) under the cutoff limts of $p:0\to \kappa$ as:
\be
N=\int_{0}^{\kappa}N(p)dp=\frac{J(v)^{2}}{9\pi^{2}}\to 0.0591729 v^2
\label{partnum_for_jinc_append}
\ee 
The energy is obtained using Eq. (\ref{energyfrombetas}) giving:
\be
E=\int_{0}^{\kappa}\hbar p N(p)dp=\frac{4J(v)^{2}\kappa}{105\pi^{2}}\to 0.0202879 v^2.
\label{energy_for_jinc_append}
\ee
Eq. (\ref{energy_for_jinc_append}) can also be confirmed in the time domain through Eq. (\ref{energyfrompower}).
\subsection{Derivation of Total Quanta, Eq.~(\ref{totalparticles}), via Frequency Space Larmor Technique} \label{appendix:N(w)}
Non-relativistically, there is a convenient method for obtaining the total particle count, using the quantum Larmor formula, Eq.~(\ref{power}):
\be
P(t) = \frac{\hbar}{6\pi} |a(t)|^2,
\ee
but in frequency space.  To start, we introduce the Fourier transform of the acceleration of the moving mirror via
\be
a(t) = \frac{1}{\sqrt{2\pi}} \int_{-\infty}^{\infty} a(\omega) e^{-i\omega t} \diff{\omega}, \ee
and its partner
\be a(\omega) = \frac{1}{\sqrt{2\pi}} \int_{-\infty}^{\infty} a(t) e^{i\omega t} \diff{t}.
\ee
According to Parseval's theorem, \(a(\omega)\) and \(a(t)\) are related by the following integral:
\be
E = \int_{-\infty}^{\infty} \frac{\hbar}{6\pi} |a(t)|^2 \diff{t} = \int_{-\infty}^{\infty} \frac{\hbar}{6\pi} |a(\omega)|^2 \diff{\omega}.
\ee
Since we want positive frequencies only, \(\int_{0}^{\infty} \cdots d\omega\) rather than \(\int_{-\infty}^{\infty} \cdots d\omega\), and the acceleration is a real function, we have:
\be
\int_{-\infty}^{\infty} |a(\omega)|^2 \diff{\omega} = 2\int_{0}^{\infty} |a(\omega)|^2 \diff{\omega}.
\ee
Thus, the total emitted radiation is:
\be
E = \int_{0}^{\infty} I(\omega) \diff{\omega} = \int_{0}^{\infty} \frac{\hbar}{3\pi} |a(\omega)|^2 \diff{\omega}, 
\ee
where the quantum energy spectrum, $I(\omega)$, is defined as
\be
I(\omega) = \frac{\hbar}{3\pi} |a(\omega)|^2.
\ee
The total quantum particle count can be understood as an integration of $I(\omega)$ over $\omega$ per quantum of energy $\hbar \omega$ accounting for a factor of 2, understanding that $\omega \to p+q$ for both frequencies: 
\be N = \int_0^\infty \frac{2 I(\omega)}{\hbar \omega} \diff{\omega} = \int_0^\infty \frac{2}{3\pi\omega}|a(\omega)|^2 \diff{\omega}.\ee
This result suggests $N(\omega) \equiv \frac{2}{3\pi\omega}|a(\omega)|^2$ is a physically valuable particle spectrum representation complementing $N(p)$ or $N(q)$. This method computes total particle count in agreement with the beta Bogolubov method, Eq.~(\ref{totalparticles}). It expresses the quantum radiation in terms of the $\omega$ frequency variable conjugate to the $t$ coordinate time variable.

\subsection{Derivation of Fourier-Bogolubov Relation Eq.~(\ref{betas}); Electron-Mirror Mapping}
\label{sec:FB_via_mapping}

First, we must derive a non-relativistic spectral distribution for an accelerating electron (charge $\mathfrak{q}$). Let us start from the Heaviside-Feynman formula for the electric field (see e.g., Eq. 23.102 of Zangwell \cite{Zangwill:1507229}):
\be
\mathbf{E} = \frac{\mathfrak{q}}{4 \pi} \mathbf{\ddot{n}}_{\text{ret}},
\ee
where we know that the spectral distribution is the Fourier transform of the unit vector that points from the retarded position of the moving charge to the observation point (see Eq. 23.103 \cite{Zangwill:1507229}):
\be
\frac{\diff{I}}{\diff{\Omega}} = \frac{\mathfrak{q}^2 r^2}{8 \pi^2} \left| \mathcal{F}\; \mathbf{\ddot{n}}_{\text{ret}} \right|^2.
\ee
We are concerned with the perpendicular component of the acceleration to the observer's line of sight for a charge moving with non-relativistic speeds:
\be
\mathbf{\ddot{n}}_{\text{ret}}  \to  \frac{a_x}{r}
= \frac{a \sin\theta}{r}\ee
This gives the spectral distribution:
\be
\frac{\diff{I}}{\diff{\Omega}} = \frac{\mathfrak{q}^2}{8 \pi^2} \sin^2\theta \left| \mathcal{F} a_t \right|^2.
\ee
Finally, considering the frequency dependence $| \mathcal{F} a_t|^2 = \omega^2|\mathcal{F} v_t|^2 $, the spectral distribution is
\be
\frac{\diff{I}}{\diff{\Omega}} = \frac{\mathfrak{q}^2\omega^2}{8 \pi^2} \sin^2 \theta \left| \mathcal{F} v_t \right|^2.
\ee
This demonstrates the non-relativistic expression of the spectral distribution from a straight-line accelerating electron as derived from the Heaviside-Feynman formula.


We may use the above non-relativistic spectral distribution in the electron-mirror mapping to derive a useful and simple formula for the quantum non-relativistic beta Bogolubov coefficients. The usual mapping uses $\omega = p+q$, and
\begin{equation}
|\beta^R_{pq}|^2 = \frac{4\pi}{\mathfrak{q}^2 \omega^2} \frac{\diff{I}}{\diff{\Omega}}(\omega,\theta), \quad \cos\theta  = \frac{p-q}{p+q}.
\label{recipe_dIdOmega_from_mirror}
\end{equation}
We need to convert from sin, rather than cos, so:
\be \sin^2\theta = \frac{4 p q}{(p+q)^2},\ee
which gives
\be
|\beta^R_{pq}|^2 = \frac{4\pi}{q^2 \omega^2}\left[\frac{q^2 \omega^2}{8\pi^2} \sin^2 \theta |v_\omega|^2\right] = \frac{1}{2\pi}\frac{4 p q}{(p+q)^2} |v_\omega|^2.\ee
To account for the total energy emitted by the moving mirror, we need to account for both sides by multiplying by a factor of 2 (this is possible due to the symmetry between $p$ and $q$). Removing the subscript $R$, gives:
\be|\beta_{pq}|^2 = \frac{4}{\pi}\frac{p q}{(p+q)^2} |v_\omega|^2.\ee
Since the Fourier identity $|v_\omega|^2 = \omega^2 |z_\omega|^2$, and $\omega = p+q$, the simple formula, Eq.~(\ref{betas}), results:
\be|\beta_{pq}|^2 = \frac{4 p q}{\pi}|z_\omega|^2.\ee
This expression directly connects Fourier analysis to the Bogolubov coefficients. It allows us to interpret the beta Bogolobov coefficient loosely as the trajectory $z(\omega)$, expressed in frequency space.

\subsection{Derivation of Fourier-Bogolubov Relation Eq.~(\ref{betas}); Mirror Model Method}
\label{sec:FB_via_mirror}

In this subsection, we derive the Fourier-Bogolubov result directly from the Bogolubov expressions without electron-mirror mapping. That is, using the relativistic Bogolubov coefficient integral, e.g., \cite{Good:2016atu}, 
\be
\beta^R_{p q} = \frac{1}{4\pi \sqrt{p q}} \int_{-\infty}^{\infty} \diff{t} \, e^{-i \omega t + i \omega_- z(t)} \left( \omega \frac{\diff{z}(t)}{\diff{t}} - \omega_- \right),\label{betaR_3}
\ee
where $\omega = p+q$ and $\omega_- = p - q$.  Let us denote the maximal speed by $s$, that is $\max(|\dot{z}|)=s$.
We want to pass to the non-relativistic limit $s \ll 1$.
To this end we introduce $\mathtt{Z}(t)$ such that
\begin{equation}
	z(t) = s \, \mathtt{Z}(t) \,, \quad
	\max\left( \abs{\dot{\mathtt{Z}}} \right) = 1 \,, \quad
	\max(|\dot{z}|)=s.
\label{Z_def}
\end{equation}
That way, we can easily isolate the $s$-scaling of the integrand. Now we write the integrand (un-normalized) of Eq.~\eqref{betaR_3} as
\begin{equation}
	I \equiv \left\{ s (p+q) \dot{\mathtt{Z}}(t) - (p-q) \right\} \, e^{-i (p+q) t } \, e^{ i s (p-q) \mathtt{Z}(t) }  \ .
\label{betaR_4}
\end{equation}
Allow us to expand Eq.~\eqref{betaR_3} via a power series in $s$:
\begin{equation}
	\beta_{p q}^R = \sum_{n=0}^\infty (\beta_{p q}^R)_n \, s^n
\label{s_expansion_1}. 
\end{equation}
The first couple of terms in Eq. \eqref{s_expansion_1} can be computed easily.  From Eq.~\eqref{betaR_4},
\begin{equation}
	(\beta_{p q}^R)_0 
		= \frac{- (p-q)}{4 \pi \sqrt{pq}} \int\limits_{-\infty}^{\infty} \diff t \, e^{-i (p+q) t } 
		= \frac{ q - p }{ 2 \sqrt{pq}} \delta( p + q ),
\label{beta_expansion_0}
\end{equation}
where we have used
\begin{equation}
	\delta(\omega)=\frac{1}{2 \pi} \int_{-\infty}^{\infty} e^{- i \omega t} d t.
\end{equation}
Since frequencies $p,q$ are both non-negative, Eq.~\eqref{beta_expansion_0} is supported only on $p = q = 0$ and does not contribute to the particle radiation.
We can therefore safely drop this term.

Let us write down the next coefficient.
Expanding the last exponential in Eq.~\eqref{betaR_4} and keeping only the terms linear in $s$ we obtain for the corresponding coefficient
\begin{equation}
	(\beta_{p q}^R)_1
		= \frac{1}{4 \pi \sqrt{pq}} \int\limits_{-\infty}^{\infty} \diff t \, \left\{ \omega \dot{\mathtt{Z}}(t) - i \omega_-^2 \mathtt{Z}(t) \right\} \, e^{-i \omega t }.
\label{beta_expansion_1}
\end{equation}
This is nothing but a Fourier transform.
To fix the notation, we normalize the Fourier transform as
\begin{equation}
	\mathtt{Z}_\omega \equiv \mathcal{F}\left[ \mathtt{Z}(t) \right](\omega) = \frac{1}{\sqrt{2\pi}} \int\limits_{-\infty}^{\infty} \diff t \, \mathtt{Z}(t) \, e^{-i \omega t } .
\label{fourier_def}
\end{equation}
Then we have
\begin{equation}
	\mathcal{F}\left[ \dot{\mathtt{Z}}(t) \right](\omega) = i \omega \mathtt{Z}_\omega .
\label{fourier_of_derivative}
\end{equation}
We obtain for Eq.~\eqref{beta_expansion_1}:
\begin{equation}
	(\beta_{p q}^R)_1
		= \frac{1}{2 \sqrt{2 \pi} \sqrt{pq}} \left\{ i \omega^2 \mathtt{Z}_{\omega} - i \omega_-^2 \mathtt{Z}_{\omega} \right\}
		= i \sqrt{\frac{2}{\pi}} \sqrt{pq} \mathtt{Z}_\omega .
\label{beta_expansion_1_1}
\end{equation}
Therefore, to the lowest non-trivial order in the slow speed expansion we can write
\begin{equation}
\begin{aligned}
	\beta_{p q}^R 
		&= s \cdot i \sqrt{\frac{2}{\pi}} \sqrt{pq} \mathtt{Z}_\omega + O(s^2), \\
		&= i \sqrt{\frac{2}{\pi}} \sqrt{pq} z_\omega + O(s^2),
\end{aligned}
\label{beta_expansion_1_2}
\end{equation}
where we used $z = s \mathtt{Z}$ from Eq.~\eqref{Z_def}.
Finally, we use $|\beta_{p q}^R|^2 = |\beta_{q p}^L|^2$ and write down the full beta (for both sides)
\begin{equation}
|\beta_{p q}|^2 
= |\beta_{p q}^R|^2 + |\beta_{p q}^L|^2 = \frac{4}{\pi} pq \abs{ z_\omega }^2 + O(s^4). \label{beta_expansion_1_3}
\end{equation}
We have recovered Eq.~\eqref{betas}, remembering $\omega = p+q$, without using the electron-mirror mapping. Note that because the $n=0$ term in the expansion Eq.~\eqref{s_expansion_1} is absent, we get $O(s^2)$ in Eq.~\eqref{beta_expansion_1_2} and we should get $O(s^3)$ in Eq.~\eqref{beta_expansion_1_3}.
However, as we will see below, Eq.~\eqref{beta_expansion_1_3} is correct up to $O(s^4)$.

\subsection{Derivation of Relativistic Corrections to the Fourier-Bogolubov relation; Eq.~(\ref{betas})}

Using this method, we can also systematically compute higher-order corrections in speed.
Indeed, let us expand the last exponential in Eq.~\eqref{betaR_4} and keep the terms proportional to $s^n$ for a fixed $n$; we obtain for the corresponding coefficient (integrand):
\begin{equation}
\begin{aligned}
	I_n
		&= \left\{ 
				\frac{i^{n-1}\omega_-^{n-1}}{(n-1)!}  \omega (\mathtt{Z}(t))^{n-1} \dot{\mathtt{Z}}(t) 
				- \frac{i^n \omega_-^{n+1}}{n!}  (\mathtt{Z}(t))^{n}
			\right\} \, e^{-i \omega t }, \\
		&= \frac{1}{n!} \left\{ 
				 i^{n-1} \omega_-^{n-1} \omega \dv{(\mathtt{Z}(t)^{n})}{t}
				- i^n \omega_-^{n+1} (\mathtt{Z}(t))^{n}
			\right\} \, e^{-i \omega t }, \\
		&= \frac{1}{n!} \left\{ 
 				i^{n} \omega_-^{n-1} \omega^2 
				- i^n \omega_-^{n+1} 
			\right\} (\mathtt{Z}(t))^{n} \, e^{-i \omega t }, \\
		&= \omega_-^{n-1}  \frac{ i^n }{n!} \left\{ 
 				\omega^2 
				- \omega_-^2 
			\right\} (\mathtt{Z}(t))^{n} \, e^{-i \omega t }.
   \end{aligned}
\end{equation}
Integrating over coordinate time and normalizing the integrand gives
\begin{equation}
\begin{aligned}
	(\beta_{p q}^R)_n	&= \frac{ \omega_-^{n-1} \sqrt{pq} }{\pi } \frac{ i^n }{n!} \int\limits_{-\infty}^{\infty} \diff t \,  (\mathtt{Z}(t))^{n} \, e^{-i \omega t }, \\
		&= \frac{ \sqrt{2} \omega_-^{n-1} \sqrt{pq} }{ \sqrt{\pi} } \frac{ i^n }{n!} \cdot \mathcal{F}\left[ \mathtt{Z}(t)^n \right](\omega).
\end{aligned}
\end{equation}
Recall our definition of the Fourier transform Eq.~\eqref{fourier_def} used in the last step.
Plugging this result into the series Eq.~\eqref{s_expansion_1} and using $z = s \mathtt{Z}$ from Eq.~\eqref{Z_def} we can write down the all-order formula for the betas (r.h.s. of the mirror):
\begin{equation}
	\beta_{p q}^R = \sqrt{ \frac{2pq}{\pi} } \sum_{n=1}^\infty \frac{ i^n \omega_-^{n-1} }{n!} \cdot \mathcal{F}\left[ z(t)^n \right](\omega),
\label{beta_all_order}
\end{equation}
recalling that the $n=0$ term, Eq.~(\ref{beta_expansion_0}), can be dropped.

By computing the absolute value square of Eq.~\eqref{beta_all_order} we can also compute the beta-squared to any desired order.
For example,
\begin{equation}
	|\beta_{p q}^R|^2 = \frac{2 p q}{\pi} \left( |z_\omega|^2 + \omega_- \Im{ z_\omega (z^2)_\omega } \right) + O(s^4).
\end{equation}
Here $(z^2)_\omega = \mathcal{F}\left[ z(t)^2 \right](\omega)$ and $\omega = p+q$.
By adding the left and right sides of the mirror we can see that the second correction term $\sim\omega_-$ cancels, which means that Eq.~\eqref{beta_expansion_1_3} is correct up to $O(s^4)$.

Using Eq.~\eqref{beta_all_order}, the next-to-leading order (NLO) term of Eq.~\eqref{beta_expansion_1_3} is
 \be ^{\textrm{NLO}}|\beta_{pq}|^2 = \frac{pq}{\pi}(p-q)^2\left(|\mathcal{F}z_t^2|^2 - \frac{4}{3} |\mathcal{F} z_t||\mathcal{F}z_t^3|\right),\label{NLO}\ee
which is confirmed with the correct NLO term in energy emission from a time-domain asymptotic integration of the relativistic power, $P = \hbar\alpha^2/6\pi$, Eq.~(\ref{power}) over coordinate time $t$. Eq.~(\ref{NLO}) term is quartic in speed $s$.

\begin{acknowledgments}
Funding comes partly from the FY2024-SGP-1-STMM Faculty Development Competitive Research Grant (FDCRGP) no.201223FD8824 and SSH20224004 at Nazarbayev University in Qazaqstan. Appreciation is given to the ROC (Taiwan) Ministry of Science and Technology (MOST), Grant no.112-2112-M-002-013, National Center for Theoretical Sciences (NCTS), and Leung Center for Cosmology and Particle Astrophysics (LeCosPA) of National Taiwan University. 
This work is also supported in part by U.S. Department of Energy Grant No. de-sc0011842.
\end{acknowledgments}


\bibliography{main} 
\end{document}